%% file: main.tex
\documentclass[submission,copyright,creativecommons]{eptcs}
\providecommand{\event}{GandALF 2024} 

\usepackage{mathtools}
\usepackage{amsthm}
\usepackage{amssymb}
\usepackage{tikz}
\usepackage{subfloat}
\usepackage{graphicx}
\usepackage{xcolor}
\usepackage{algorithm}
\usepackage{microtype}
\usepackage[noend]{algpseudocode}
\usepackage{xspace}
\usepackage{url}

\usepackage{iftex}

\ifpdf
  \usepackage{underscore}         
  \usepackage[T1]{fontenc}        
\else
  \usepackage{breakurl}           
\fi
\usetikzlibrary{automata,positioning}

\newcommand{\Nat}{\mathbb{N}}
\newcommand{\Lang}{\mathcal{L}}

\newcommand{\mfalse}{\mathtt{false}}
\newcommand{\mtrue}{\mathtt{true}}
\newcommand{\Fib}{\mathtt{Fib}} 

\def\CC{{C\nolinebreak[4]\hspace{-.05em}\raisebox{.4ex}{\tiny\bf ++}}\xspace}

\algrenewcommand\algorithmicrequire{\textbf{Input:}}
\algrenewcommand\algorithmicensure{\textbf{Output:}}

\algblockdefx[PARNAME]{ParDo}{EndParDo}%
[1]{\textbf{do in parallel for } #1}%
{}
\makeatletter
\ifthenelse{\equal{\ALG@noend}{t}}%
  {\algtext*{EndParDo}}
  {}%
\makeatother

\theoremstyle{definition}
\newtheorem{theorem}{Theorem}
\newtheorem{definition}[theorem]{Definition}

\title{An Evaluation of Massively Parallel Algorithms for DFA Minimization}

\author{Jan Martens
\institute{Leiden University\\
The Netherlands} 
\email{j.j.m.martens@liacs.leidenuniv.nl}
 \and Anton Wijs
 \institute{Eindhoven University of Technology\\
The Netherlands} 
\email{a.j.wijs@tue.nl}
}

\newcommand{\titlerunning}{An Evaluation of Massively Parallel Algorithms for DFA Minimization}
\newcommand{\authorrunning}{J.J.M. Martens \& A.J. Wijs}

\hypersetup{
  bookmarksnumbered,
  pdftitle    = {\titlerunning},
  pdfauthor   = {\authorrunning},
  pdfsubject  = {EPTCS},               
  pdfkeywords = {dfa minimization, gpu, parallel minimization} 
}

\begin{document}
\maketitle
\begin{abstract}
We study parallel algorithms for the minimization of Deterministic Finite
Automata~(DFAs). In particular, we implement four different massively parallel
algorithms for DFA minimization on Graphics Processing Units~(GPUs). Our results
confirm the expectations that the algorithm with the theoretically best time
complexity is not practically suitable to run on GPUs due to the large amount of
resources needed. We empirically verify that parallel partition refinement
algorithms from the literature perform better in practice, even though their
time complexity is worse. Lastly, we introduce a novel algorithm based on
partition refinement with an extra parallel partial transitive closure step and
show that on specific benchmarks it has better run-time complexity and performs
better in practice.    
\end{abstract}

\section{Introduction}
In contrast to sequential chips, the processing power of parallel devices keeps
increasing. Graphics Processing Units, or GPUs, are examples of such devices.
Originating from the need to do simple computations for many (independent)
pixels to generate graphics, GPUs have also shown useful as computational
powerhouses, and led to general-purpose computing on GPUs~(GPGPU). Most
convincingly, GPUs have become indispensable in training models for artificial
intelligence. Because of the enormous potential of GPUs, it is important to
investigate how computational problem solving can be accelerated with them.

Deterministic Finite Automata~(DFAs) are one of the simplest computational
formalisms. The natural problem of computing a minimal machine that is
equivalent to a given machine w.r.t.\ the input is omnipresent in the field of theoretical computer
science. In the case of DFAs the problem has a rich history. The first method
that computes a minimal DFA dates back to Moore's
framework~\cite{moore1956gedanken}, and is a \emph{partition refinement}
algorithm. Later, this algorithm was adapted by
Hopcroft~\cite{hopcroft1971DFAmin} to a quasi-linear time algorithm.

The complexity class known as Nick's Class~($NC$) consists of the problems that
can be solved in polylogarithmic time with a parallel
machine using a polynomial number of parallel processors. It is an open question
whether \smash{$\textit{NC} \stackrel{?}{=} \textit{P}$}, but it is widely believed that this is not the case.
Similar to the assumption that
 decision problems not in $\textit{P}$
  are inherently
difficult~(known as Cobham's thesis), we can think of $P$-complete problems as
being inherently sequential. 

The problem of minimizing DFAs is known to be in
\textit{NC}~\cite{cho1992parallelcoarsestset}, which intuitively means it can be
efficiently computed in parallel. In contrast, the problem of computing
bisimilarity on non-deterministic structures is known to be
$P$-complete~\cite{balcazar1992}. Interestingly, the most efficient sequential
algorithms for these two problems, i.e., Hopcroft's
algorithm~\cite{hopcroft1971DFAmin} and an algorithm based on
Paige-Tarjan~\cite{paige1987three}, respectively, are very similar. In
particular, these algorithms are both \emph{partition refinement} algorithms.  


The parallel algorithms studied for computing 
bisimilarity on non-deterministic structures 
 and DFA minimization are also partition refinement
algorithms~\cite{martens2022linear,ravikumar1996paralleldfa,
tewari2002paralleldfa,wijs2015}. Since DFA minimization is in $\textit{NC}$,
there is a parallel sublinear time algorithm. However, none of these partition
refinement algorithms studied have a sublinear run-time. A linear lower bound
for the parallel run-time was proven in~\cite{groote2023lowerbounds} for any
parallel partition refinement algorithm deciding bisimilarity, and this result
also directly applies to deterministic structures such as DFA minimization. This
means that no partition refinement algorithm can achieve the theoretically
optimal run-time on parallel machines.  
It is therefore interesting to investigate whether there is an algorithm that is not a
partition refinement algorithm that performs better in parallel than partition refinement algorithms.  
  
The algorithm introduced in~\cite{cho1992parallelcoarsestset} runs in
logarithmic time. However, the work is mainly theoretical and the large amount
of parallel processors and memory required makes it unlikely to scale well in
practice. The main constraint here is the need to compute the transitive closure for
the underlying graph of the DFA. It seems hard to find a significant improvement
in the number of parallel processors needed. 

In this paper we compare implementations of different parallel algorithms
for DFA minimization on GPUs, using the various parallel algorithms proposed in the literature
as a basis. We establish that the logarithmic runtime
complexity with the construction from~\cite{cho1992parallelcoarsestset} is not
feasible due to the large amount of processors needed. Additionally, we find
that on our benchmarks the partition refinement algorithm that uses sorting in
each iteration performs better than the naive splitting strategy on the more
diverse benchmarks from the VLTS benchmark set,\footnote{https://cadp.inria.fr/resources/vlts (visited on: 19-04-2024).} but worse for benchmarks that are
known to be hard for partition refinement algorithms. Finally, we show a method
of adding a partial transitive closure as a preprocessing step that can
significantly increase the performance on benchmarks with a very specific shape.

\section{Preliminaries}
We write $\mathbb{B} = \{\mtrue, \mfalse\}$ for the set of booleans, $\Nat$ for the set
of natural numbers, and for numbers $i,j\in \Nat$ we define $[i,j] = \{ c \in
\Nat \mid i \leq c \leq j\} \subseteq \Nat$, the closed interval from $i$ to
$j$. Given an alphabet $\Sigma$, a sequence $a_1 a_2 \ldots a_n$ of symbols from
$\Sigma$ is called a word. We write $\Sigma^*$ for the set containing all finite
sequences of letters in $\Sigma$. The empty-sequence consisting of no symbols is
written as $\varepsilon$. 

\begin{definition}{(Deterministic Finite Automaton)}
  A deterministic finite automaton~(DFA) $A = (Q, \Sigma, \delta, F, q_0)$ is a five-tuple consisting of:
  \begin{itemize}\setlength\itemsep{0em}
    \item a finite set of states $Q$,
    \item a finite alphabet $\Sigma$, 
    \item a transition function $\delta: Q\times \Sigma \to Q$, 
    \item a set of accepting states $F\subseteq Q$, and
    \item an initial state $q_0 \in Q$. 
  \end{itemize}
\end{definition}

We sometimes write $q \xrightarrow{a} q'$ if $\delta(q, a) = q'$. The function
$\delta^*: Q \times \Sigma^* \to Q$ extends the transition function to words and
is defined inductively for all words in $\Sigma^*$ as follows:
\begin{align*}
  \delta^*(q,\varepsilon) &= q \\
  \delta^*(q,aw) &= \delta^*(\delta(q,a), w)
\end{align*}

Given a DFA $A=(Q,\Sigma,\delta, F,q_0)$, a word $w \in \Sigma^*$ is
 \emph{accepted} iff $\delta^*(q_0, w) \in F$. The language of a DFA $A$,
 notation $\Lang(A)$, is the set of all words $w \in \Sigma^*$ that are accepted
 by $A$. 

We consider the problem of computing the minimal DFA, i.e., given a DFA $A$,
identifying the DFA $A'$ with the smallest number of states such that $\Lang(A')
= \Lang(A)$. 

Minimizing DFAs consists of combining undistinguishable states and deleting
unreachable states. The main part of the problem consists of combining states, and
removing unreachable states can be seen as a simple pre-processsing step. For
the remainder of the paper, we assume that all the states in a DFA are reachable
from its initial state. The
algorithms can be seen as computing bisimilarity, or the coarsest set partition
problem, without the preprocessing step that removes unreachable states.

\vspace{-8pt}

\paragraph*{Representation.} For an input
automaton $A= (Q, \Sigma, \delta, F, q_0)$, we assume that the states in $Q$ and letters in
$\Sigma$ are represented by unique indices, i.e., $Q = \{0, \dots, \lvert Q
\rvert\}$ and $\Sigma = \{0, \dots \lvert \Sigma \rvert\}$. The transition
function $\delta$ is represented by $\lvert \Sigma \rvert$ arrays of length
$\lvert Q \rvert$, such that for state $q\in Q$ and letter $a\in \Sigma$, $\delta[a][q] = \delta(q,a)$.  

\vspace{-8pt}

\paragraph*{The PRAM Model.} The complexities we mention assume the model of
the \emph{Parallel Random Access Machine~(PRAM)}. The PRAM is a natural extension
of the RAM model, where parallel processors have access to a shared memory. A PRAM consists of a sequence of processors 
$P_0, P_1, \dots$ and a function $\mathcal{P}$ that given the size of the input defines a bound on the number of processors used. 

Each processor $P_i$ has the natural instructions of a normal RAM and in addition has an instruction to
retrieve its unique index $i$. All processors run the same program in lock-step, using their index to identify the data they need to access.
This parallel processing is called single-instruction multiple data~(SIMD).  

There are many different ways to handle data-races. 
We assume the concurrent read, concurrent write~(CRCW) model following~\cite{stockmeyer1984simulation}, 
where processors are allowed to read from and write to the same memory location concurrently.
After multiple concurrent writes to the same memory location, that location contains the result
of one of those writes.

\vspace{-8pt}

\paragraph*{GPUs.}
While in reality, no device completely adheres to the PRAM model, recent hardware advancements has led
to devices that are getting better and better at approximating this model. The GPU, in particular, is a very suitable
target platform for PRAM algorithms, as it has been specifically designed for SIMD processing.
The performance of GPU programs typically relies on launching tens to hundreds
of thousands of threads, as the performance of these programs is often memory-bound: accessing the input data
in the GPU's \emph{global} memory, in NVIDIA CUDA terminology, is relatively slow. This latency can be hidden
by a GPU via fast context switching between threads. As one thread is waiting for data to be retrieved, another thread
is executed in the meantime on the same processor. It is this fast context switching between threads that allows GPUs,
typically equipped with several thousands of cores, to virtually execute hundreds of thousands of threads concurrently.
In the current work, we employ NVIDIA GPUs, programs for which can be written in CUDA \CC.

\section{The algorithms}
\subsection{Transitive closure} 
The first algorithm we discuss has theoretical polylogarithmic
runtime~\cite{cho1992parallelcoarsestset}. However, the large amount of memory
and parallel processors it uses makes it unlikely to work in practice. Here we
confirm this fact.  

The idea is rather simple; build a graph with nodes $V = Q\times Q$ and edges
$E$ containing $(q, q') \rightarrow (p, p')$ iff there is a letter $a\in \Sigma$
such that $\delta(q,a) = p$ and $\delta(q', a) = p'$. Initially, in the array
$\mathtt{Apart}$ we label the nodes $(q,q') \in V$ to be inequivalent if $q_1
\in F \iff q_2 \not \in F$. Any two states $q_1, q_2 \in Q$ are not equivalent
iff they were initially labelled in $\mathtt{Apart}$ or there is a path to a
labelled node. Computing this reachability of false nodes can be seen as
computing the transitive closure in the directed graph $(V,E$) containing $n^2$
nodes. In parallel this computation can be done in polylogarithmic running time
using $O(\lvert V\rvert^3)$ parallel processors~\cite[Chapter 5.5.]{jaja1992introduction}. 


Algorithm~\ref{alg:fulltrans}, which we refer to as \texttt{trans},
 implements this idea. First, at lines~4--7 (l.4--7), the
graph is constructed, inequivalent nodes labelled in the $\mathit{Apart}$ data
structure and the edges stored in the adjacency matrix $\mathit{Reach}$. Next,
the parallel transitive closure of $\mathit{Reach}$ is computed and
$\mathit{Apart}$ updated accordingly. If in an iteration there is no new pair of
states labelled $\mathit{Apart}$ the algorithm is finished. The minimal automaton
is represented in the graph where states $q, q' \in Q$ can be combined if $\neg
\mathit{Apart}(q,q')$.  

Computing the transitive closure for a directed graph in logarithmic time
requires many processors. Our naive implementation requires $\lvert V\rvert^3$
parallel processors. Given a DFA with $n$ states, this means that since
$\lvert V \rvert = n^2$, we require $n^6$ processors. Theoretically, more
efficient methods are known for computing the transitive closure, which uses
matrix multiplication. Matrix multiplication can be computed with
$O(n^\omega)$ operations, where currently $\omega \leq 2.372\dots$, this
means we can compute our transitive closure with $O(\lvert V \rvert^\omega)$
processors. Since these algorithms are non-trivial and already $\lvert V \rvert
= n^2$, we believe these improvements would not significantly change the results
mentioned here.


\begin{algorithm}[tb]
  \begin{algorithmic}[1]
    \Require{A DFA $A = (Q, \Sigma, \delta, F, q_0)$ where $\lvert Q \rvert = n$}
    \Ensure{The minimal quotient automaton represented in the matrix $\mathit{Apart}$}
    \State $V :: Q\times Q$ \Comment{Nodes of graph consisting of pair of states}
    \State $\mathit{Apart} ::$ Array$[n^2]$ of type $\mathbb{B}$ 
    \State $\mathit{Reach} ::$ Array$[n^2][n^2]$ of type $\mathbb{B}$ \Comment{Represents reachability in $V$, initially $\mfalse$}
    \ParDo{$(q, q')\in V$ }
      \Comment{Initializes data structures in parallel.}
      \State $\mathit{Apart}[(q,q')] := (q \in F \iff q'\not\in F)$ \Comment{State initially unequal}
      \ForAll{$a\in \Sigma$}
        \State $\mathit{Reach}[(q, q')][(\delta(q,a), \delta(q',a))] := \mtrue$
      \EndFor
    \EndParDo
    \State $\mathit{stable} := \mfalse$
    \While{$\neg \mathit{stable}$} 
      \State $\mathit{stable} := \mtrue$
       \ParDo {$s,t,u \in V$ }
        \If{$\mathit{Reach}[s][t] \text{ and } \mathit{Reach}[t][u] \text{ and } \neg \mathit{Reach}[s][u]$}
          \State $\mathit{Reach}[s][u] := \mtrue$
        \EndIf
        \EndParDo
      \ParDo {$s,t \in V$}
        \If{$\mathit{Reach}[s][t] \text{ and } \mathit{Apart}[t] \text{ and } \neg \mathit{Apart}[s]$}
          \State $\mathit{Apart}[s] := \mtrue$
          \State $\mathit{stable} := \mfalse$
        \EndIf
      \EndParDo
    \EndWhile 
\end{algorithmic}
\caption{\label{alg:fulltrans} Transitive DFA minimization~$\mathtt{trans}$.}
\end{algorithm}

\subsection{Naive partition refinement}
The next algorithm, \texttt{naivePR},
 is an adaptation of the parallel algorithm for bisimilarity
checking of Kripke structures from~\cite{martens2022linear}.
The program runs on a PRAM with $\max(n, m)$ processes, where $n$ is the number
of states and $m = \lvert \Sigma \rvert * n$ is the number of transitions in the
input DFA.
 
The algorithm applies \emph{partition refinement}: states are initially partitioned into \emph{blocks},
and the algorithm repeatedly splits blocks into smaller blocks until a fix-point is reached.
Once this has happened, each block represents one state of the minimized DFA.

When splitting blocks in parallel, one particular challenge is how to identify
newly created blocks, as each new block requires a unique identifier.
Algorithm~\ref{alg:partref} does this by means of a leader election procedure:
for each block, one of its states is elected leader, meaning that it is used as
an identifier to refer to the block. In this way each iteration of the algorithm
takes constant time if performed on a parallel machine that has concurrent
writes.

In Algorithm~\ref{alg:partref}, at l.\ref{l:partref-block}, an array
\textit{block} is initialized that defines for every state in $Q$ its current
block (as represented by a leader in $Q$). An array \textit{new\_leader} is
defined at l.\ref{l:partref-newleader} that is used to elect new leaders. At
l.\ref{l:partref-initleaders}, the initial leaders are selected: one state $q_f
\in F$ for the block consisting of all the accepting states $q \in F$, and one
state $q_n \in Q \setminus F$ for all the non-accepting states $q' \in Q
\setminus F$. The array \textit{block} is subsequently initialised using these
leaders (l.\ref{l:partref-for-initblock}--\ref{l:partref-initblock}).

Next, partition refinement is applied inside the \textbf{while}-loop at l.\ref{l:partref-whileneqstable}. The variable
\textit{stable} is used to monitor whether a fix-point has been reached, which has happened as soon as no blocks
can be split any further. At the start of each iteration through the \textbf{while}-loop, \textit{stable} is set to
$\mtrue$ (l.\ref{l:partref-setstable}). Next, all transitions of the DFA are processed in parallel (l.\ref{l:partref-leaderfor}),
and for each transition $q \xrightarrow{a} q'$, it is checked whether $\mathit{block}[q']$ differs from the block that
the leader $\mathit{block}[q]$ can reach via an $a$-transition. If it does, then $q$ should be separated from its
leader. At l.\ref{l:partref-leaderelect}, $q$ is assigned to $\mathit{new\_leader}[\mathit{block}[q]]$, the latter being the
position where the leader for the new block will be elected. Here, the result of concurrent writes, as allowed by the
PRAM CRCW model, is used for leader election.

Subsequently, when l.\ref{l:partref-getleaderfor} is reached, \textit{new\_leader} contains the newly elected leaders:
specifically, at $\mathit{new\_leader}[\mathit{block}[q]]$, the leader for the new block created by splitting off states from
$\mathit{block}[q]$ is stored. In the parallel loop of l.\ref{l:partref-getleaderfor}, the transitions are once more processed
in parallel, and whenever a state turns out to differ from its leader regarding block reachability (l.\ref{l:partref-get-leaderdiff}),
the leader of that state is updated (l.\ref{l:partref-getnewleader}). Finally, since a block has been split, \textit{stable}
is set to $\mfalse$ at l.\ref{l:partref-notstable}.
 
The largest difference between Algorithm~\ref{alg:partref} and the original
algorithm~\cite{martens2022linear} is that Algorithm~\ref{alg:partref} splits
blocks directly w.r.t.\ the leader, as opposed to first selecting one particular
block as \emph{splitter}, and splitting those blocks in which some states differ
w.r.t.\ their leader concerning the ability to reach the splitter. The reason
for this difference is that for DFAs, comparing the outgoing transitions of two
states is much more straightforward, as each state has exactly one outgoing
transition for every $a \in \Sigma$. In the setting of LTSs, due to
non-determinism it is not possible to directly compare the behaviour of a state
with the leader state, and hence a fixed splitter is chosen before. 

\begin{algorithm}
  \begin{algorithmic}[1]
    \Require{A DFA $A = (Q, \Sigma, \delta, F, q_0)$ where $\lvert Q \rvert = n$}
    \Ensure{The minimal quotient automaton represented in the array $block$}
    \State $\mathit{block} ::$ Array$[n]$ of type $Q$ \label{l:partref-block}
    \State $\mathit{new\_leader} ::$ Array$[n]$ of type $Q$ \label{l:partref-newleader}
    \State Select initial leader states~$q_f \in F$ and $q_{n}\in Q\setminus F$ \label{l:partref-initleaders}
    \ParDo{$q\in Q$} \label{l:partref-for-initblock}
      \State $\mathit{block}[q] := (q \in F\ ?\ q_f : q_n)$ \Comment Initialize \label{l:partref-initblock}
    \EndParDo
    \State $\mathit{stable} := \mfalse$
    \While{$\neg \mathit{stable}$} \label{l:partref-whileneqstable}
      \State $\mathit{stable} := \mtrue$ \label{l:partref-setstable}
      \ParDo{$q,a\in Q \times \Sigma$} \label{l:partref-leaderfor}
        \If{$\mathit{block}[\delta(q,a)] \neq \mathit{block}[\delta(\mathit{block}[q],a)]$} \label{l:partref-leaderdiff}
          \State $\mathit{new\_leader[block[q]]} := q$ \Comment Leader election \label{l:partref-leaderelect}
        \EndIf
      \EndParDo
      \ParDo{$q, a \in Q \times \Sigma$} \label{l:partref-getleaderfor}
        \If{$\mathit{block}[\delta(q,a)] \neq \mathit{block}[\delta(\mathit{block}[q],a)]$} \label{l:partref-get-leaderdiff}
          \State $\mathit{block[q]} := \mathit{new\_leader}[{\mathit{block}[q]}]$ \Comment Split from leader \label{l:partref-getnewleader}
          \State $\mathit{stable} := \mfalse$ \label{l:partref-notstable}
        \EndIf
      \EndParDo
    \EndWhile
\end{algorithmic}
\caption{\label{alg:partref} Parallel leader-election-based algorithm~$\mathtt{naivePR}$.}
\end{algorithm}

\begin{algorithm}
  \begin{algorithmic}[1]
    \Require{A DFA $A = (Q, \Sigma, \delta, F, q_0)$, where $\lvert Q \rvert = n$} 
    \Ensure{The minimal quotient automaton represented in the array $block$}
    \State $\mathit{block} ::$ Array$[n]$ of type $Q$ 
    \State $\mathit{new\_block} :: Q$
    \State $\mathit{leader} :: Q$
    \State $\mathit{new\_leader} ::$ Array$[n]$ of type $Q$
    \State Select initial leader states~$q_f \in F$ and $q_{n}\in Q\setminus F$
    \ParDo{$q\in Q$}
      \State $\mathit{new\_leader}[q] := \bot$ \Comment Initialize
      \State $\mathit{block}[q] := (q \in F\ ?\ q_f : q_n)$
    \EndParDo
    \While{$\neg \mathit{stable}$}
      \State $\mathit{stable} := \mtrue$
      \ParDo{$q,a\in Q \times \Sigma$} \label{l:partref-cas-leaderfor}
        \State $\mathit{leader} := \mathit{block}[q]$
        \If{$\mathit{block}[\delta(q,a)] \neq \mathit{block}[\delta(\mathit{leader},a)]$}
          \State $\{\mathit{new\_block} := \mathit{new\_leader[leader]};$ \Comment Leader election with CAS \label{l:partref-cas-casbegin}
          \State $\mathit{new\_leader[leader]} = \bot\ ?\ new\_leader[leader] := q\}$ \label{l:partref-cas-casend}
          \State $\mathit{block[q]} := \mathit{new\_block} = \bot\ ?\ q: \mathit{new\_block}$ \Comment Split from leader \label{l:partref-cas-split}
          \State $\mathit{stable} := \mfalse$ \label{l:partref-cas-notstable}
        \EndIf
      \EndParDo
      \ParDo{$q\in Q$} \label{l:partref-cas-resetbegin}
        \State $\mathit{new\_leader[q]} := \bot$ \label{l:partref-cas-resetend}
      \EndParDo
    \EndWhile
\end{algorithmic}
\caption{\label{alg:partref-cas} Parallel leader-election based $\mathtt{naivePR}$ with atomics.}
\end{algorithm}

In Algorithm~\ref{alg:partref}, leader election is performed in two phases: in the first phase (l.\ref{l:partref-leaderfor}--\ref{l:partref-leaderelect}),
states are written to $\mathit{new\_leader}$ to elect new leaders, and the results are subsequently read at l.\ref{l:partref-getleaderfor}--\ref{l:partref-notstable}.
One could argue that this is inefficient, and that it would perhaps be better to combine these two phases. This is possible, but it requires the use of
atomic \textit{compare-and-swap}~(CAS) operations. This is illustrated in Algorithm~\ref{alg:partref-cas}.
In the single loop at l.\ref{l:partref-cas-leaderfor}--\ref{l:partref-cas-notstable}, new leaders are written to \emph{and} read from
$\mathit{new\_leader}$. At l.\ref{l:partref-cas-casbegin}--\ref{l:partref-cas-casend}, the use of a compare-and-swap operation
is described: in one atomic operation, the current value stored at $\mathit{new\_leader[leader]}$ is stored in $\textit{new\_block}$, and
it is checked whether $\mathit{new\_leader[leader]}$ is equal to the initial value $\bot$, and if it is, it is set to $q$.
Next, at l.\ref{l:partref-cas-split}, if \textit{new\_block} is equal to $\bot$, it means $q$ has been elected as leader. Otherwise,
\textit{new\_block} indicates which state is the new leader. Note that for this to work, after execution of the loop at l.\ref{l:partref-cas-leaderfor},
the values of \textit{new\_leader} have to be reset to $\bot$.

In practice, we experienced that a GPU implementation (in CUDA 12.2) of Algorithm~\ref{alg:partref-cas} exhibits similar runtimes compared to a GPU implementation
of Algorithm~\ref{alg:partref}. The benefit of merging the loops seems to be negated by the use of atomic operations. For this reason, when discussing
the experiments in Section~\ref{sec:results}, we do not involve Algorithm~\ref{alg:partref-cas}.

\subsection{Sorting arrays}
The next algorithm is an algorithm inspired by~\cite{ravikumar1996paralleldfa,
tewari2002paralleldfa}. Similar to Algorithm~\ref{alg:partref}, this algorithm
also performs partition refinement, but instead of doing so using leader
elections, it repeatedly computes a \emph{signature} for every state, and sorts
the states w.r.t.\ their signatures. This method allows splitting a block in
more than two subblocks with the downside that each iteration takes more than
constant parallel time. 

The algorithm from~\cite{tewari2002paralleldfa} uses hashing to construct and
compare signatures. Since sorting arrays is a very native operation on GPUs, we
follow~\cite{ravikumar1996paralleldfa} and use a sorting approach to construct
the new blocks.

Algorithm~\ref{alg:sort} presents this approach as \texttt{sortPR}. Again, an array \textit{block} is created (l.\ref{l:sort-block}).
In addition, an array \textit{state} is used for sorting the states (l.\ref{l:sort-state}). The signature of
a state consists of a list of block IDs, one for each $a \in \Sigma$: $\mathit{signature}[q][a]$ is equal to $q'$
iff $q \xrightarrow{a} q''$ and $\mathit{block}[q''] = q'$.

The array \textit{new\_block} is used to store the results of assigning new blocks to states (l.\ref{l:sort-newblock}).
Finally, the current number of blocks is stored at l.\ref{l:sort-numblocks} in \textit{num\_blocks}.

Next, at l.\ref{l:sort-for-initblockstate}--\ref{l:sort-initstate}, \textit{block} and \textit{state} are initialised.
The block consisting of all accepting states is given ID 0, while the other states are assigned to block $1$
(l.\ref{l:sort-initblock}). All the states are added to \textit{state} at l.\ref{l:sort-initstate}.

In the loop at l.\ref{l:sort-repeat}--\ref{l:sort-until}, the partition
refinement is performed until a fix-point has been reached, i.e., the number of
blocks has not increased (l.\ref{l:sort-until}). In each iteration of this loop,
the following is performed. First, in parallel, the signatures are updated
(l.\ref{l:sort-for-makesigs}--\ref{l:sort-makesigs}). After that, \textit{state}
is sorted in parallel, using \textit{signature} to compare states. The
comparison function is given at
l.\ref{l:sort-compare-begin}--\ref{l:sort-compare-end}. First, states are
compared based on the block they reside in. If they reside in the same block,
then the blocks they can reach via outgoing transitions are compared. Note
that the for loop starting in (l.\ref{l:sort-compare-loop}) is sequential and
requires iterating over the alphabet letters in a fixed order.

Once \textit{state} has been sorted, the parallel \emph{adjacent difference} is computed and stored in
\textit{new\_block}. The result of this is that $\mathit{new\_block}[0] = \mathit{state}[0]$ and for all
$0 < i < n$, $\mathit{new\_block}[i] = \mathit{are\_neq}(\mathit{state}[i], \mathit{state}[i-1])$, with \textit{are\_neq}
as defined at l.\ref{l:sort-areneq-begin}--\ref{l:sort-areneq-end}.
Once $\mathit{new\_block}[0]$ has been reset to $0$ (l.\ref{l:sort-set0}), \textit{new\_block} contains only $0$'s
and $1$'s, with each $1$ identifying the start of a new block. At l.\ref{l:sort-scan}, an \emph{inclusive scan} is
performed in parallel, resulting in \textit{new\_block} having been updated in such a way that for each $0 \leq i < n$,
$\mathit{new\_block}[i] = \sum_{0 \leq j \leq i} \mathit{new\_block}'[j]$, with \smash{$\mathit{new\_block}'$} referring
to \textit{new\_block} at the start of executing l.\ref{l:sort-scan}.

Now, for all $0 \leq i < n$, $\mathit{new\_block}[i]$ contains the new block of state $\mathit{state}[i]$.
At l.\ref{l:sort-for-updateblock}--\ref{l:sort-updateblock}, \textit{block} is updated in parallel to reflect this.
As the largest new block ID can be found at $\mathit{new\_block}[n-1]$, this location can be used to determine
the new number of blocks at l.\ref{l:sort-until}.

\begin{algorithm}
\begin{algorithmic}[1]
  \Require{A DFA $A = (Q, \Sigma, \delta, F, q_0)$, where $\lvert Q \rvert = n$ and $\lvert \Sigma \rvert = k$} 
  \Ensure{The minimal quotient automaton represented in the array $block$}

  \State $\mathit{block} ::$ Array$[n]$ of type $\Nat$ \label{l:sort-block}
  \State $\mathit{state} ::$ Array$[n]$ of type $Q$ \label{l:sort-state}
  \State $\mathit{signature} :: $ Array$[n][k]$ of type $Q$ \label{l:sort-signature}
  \State $\mathit{new\_block} :: $ Array$[n]$ of type $\Nat$ \label{l:sort-newblock}
  \State $\mathit{num\_blocks} := 2$ \label{l:sort-numblocks}
    \ParDo{$q\in Q$} \label{l:sort-for-initblockstate}
      \State $\mathit{block}[q] := (q \in F\ ?\ 0 : 1)$ \Comment Initialize \label{l:sort-initblock}
      \State $\mathit{state}[q] := q$ \label{l:sort-initstate}
    \EndParDo
  \Repeat \label{l:sort-repeat}
    \State $\mathit{num\_blocks} := \mathit{new\_block}[n-1] + 1$ \Comment Number of blocks before iteration
    \ParDo{$q, a \in Q \times \Sigma$} \label{l:sort-for-makesigs}
        \State $\mathit{signature}[q][a] := block[\delta(q, a)]$ \label{l:sort-makesigs}
    \EndParDo
      \State $\mathtt{sort}(\mathit{state}, \textsc{compare})$ \label{l:sort-sortstates}
      \State $\mathit{new\_block} := \mathtt{adjacent\_diff}(\mathit{state}, \textsc{are\_neq})$ \Comment Place $1$ for each change \label{l:sort-diff}
      \State $\mathit{new\_block}[0] := 0$ \label{l:sort-set0}
      \State $\mathit{new\_block} := \mathtt{inclusive\_scan}(\mathit{new\_block})$ \Comment Compute new block labels \label{l:sort-scan}
      \ParDo{$q\in Q$} \label{l:sort-for-updateblock}
        \State $\textit{block}[state[q]] = \textit{new\_block}[q]$ \label{l:sort-updateblock}
      \EndParDo
  \Until{$\mathit{new\_block[n-1]} + 1 = \mathit{num\_blocks}$} \label{l:sort-until}
\vspace{0.4cm}
\Function{compare}{$q_1$, $q_2$} \label{l:sort-compare-begin}
\If{$\mathit{block}[q_1] > \mathit{block}[q_2]$}
	\Return $\mfalse$
\EndIf
\If{$\mathit{block}[q_1] < \mathit{block}[q_2]$}
	\Return $\mtrue$
\EndIf
\ForAll{$a \in \Sigma$} \label{l:sort-compare-loop}
	\If{$\mathit{signature}[q_1][a] > \mathit{signature}[q_2][a]$}
		\Return $\mfalse$
	\EndIf
	\If{$\mathit{signature}[q_1][a] < \mathit{signature}[q_2][a]$}
		\Return $\mtrue$
	\EndIf
\EndFor
\State \Return $\mfalse$
\EndFunction \label{l:sort-compare-end}
\vspace{0.4cm}
\Function{are\_neq}{$q_1$, $q_2$} \label{l:sort-areneq-begin}
\If{$\mathit{block}[q_1] \neq \mathit{block}[q_2]$}
	\Return $\mtrue$
\EndIf
\ForAll{$a \in \Sigma$}
	\If{$\mathit{signature}[q_1][a] \neq \mathit{signature}[q_2][a]$}
		\Return $\mtrue$
	\EndIf
\EndFor
\State \Return $\mfalse$
\EndFunction \label{l:sort-areneq-end}
\end{algorithmic}
\caption{\label{alg:sort} Parallel sorting-based algorithm~$\mathtt{sortPR}$}
\end{algorithm}


In~\cite{ravikumar1996paralleldfa} it is shown that on average this algorithm
has polylogarithmic run-time complexity. The argument given uses the fact that
on uniformly sampled DFAs almost all pairs of states have a shortest
distinguishing word of polylogarithmic depth. This fact is attributed
to~\cite{trachtenbrot1973finite}. Although this is true for uniformly sampled
DFAs, we like to stress that for many use cases and real-life applications this
bound does not apply. For example, in the Fibonacci automata presented in
Section~\ref{sec:results} this is not the case. In the automaton $\Fib_i$
containing $n$ states, there is a pair of states for which the shortest
distinguishing word, and thus also the number of iterations, has length $n{-}2$.

\subsection{Partition refinement using partial transitive closure}
In this section, we present a new algorithm \texttt{transPR}. The main idea of
the algorithm is to perform partition refinement like the algorithms before, 
but in the initialization compute the transitive closure on each distinct
 letter. After this initialization step, we use $\texttt{naivePR}$ to
 complete the minimization.
 
 This approach is presented in Algorithm~\ref{alg:partref-trans}. This
 is done in a data-parallel way which is also used for prefix sum and finding
 the end of a linked list~\cite{Hillis86}. On some DFAs, with a rather specific
 structure, this method exponentially improves the runtime compared to the
 other partition refinement algorithms.

\begin{algorithm}
  \begin{algorithmic}[1]
    \State $\Sigma^T := \{a^{2^i} \mid a\in \Sigma, i \in [0, \lfloor\log n \rfloor]\}$
    \State $\delta^T :: Q \times \Sigma^T \mapsto Q$ 
    \State $\delta^T(q, a) := \delta(q,a)$ for all $a\in \Sigma$
    \ForAll{$i \in [1, \lfloor\log n \rfloor]$}
      \ForAll{$a\in \Sigma$}
        \ParDo{$q \in Q$}
          \State $\delta^T(q, a^{2^i}) := \delta^T(\delta^T(q, a^{2^{i-1}}),a^{2^{i-1}})$\label{l:log-trans-closure}
        \EndParDo
      \EndFor
    \EndFor
    \State Perform $\texttt{naivePR}$ on the DFA $A' = (Q, \Sigma^T, \delta^T, F, q_0)$
  \end{algorithmic}
  \caption{\label{alg:partref-trans} Parallel partition refinement with transitive closure~$\mathtt{transPR}$}
  \end{algorithm}

The algorithm works by adding letters for increasingly large words of the same
letters. Given an input DFA $A= (Q, \Sigma, \delta, F, q_0)$, we construct a DFA
$A' = (Q, \Sigma^T, \delta^T, F, q_0)$ which has the same set of states and
final states, but a larger alphabet $\Sigma^T$. The alphabet contains the
letters $a^{2^0}, a^{2^1}, \dots, a^{2^{\lfloor \log n\rfloor}}$ for each
original letter $a\in \Sigma$. The transition function is computed such that for
each new symbol $a^k\in \Sigma^T$ the transition function $\delta^T(q_1,a^k) =
q_k$ if in the original DFA $Q$ there are states  $q_1, \dots q_k \in Q$ such
that $\delta(q_{i}, a) = q_{i+1}$ for each $i\in[1,k]$. This can be computed in
a logarithmic number of parallel steps, by using the previously computed
transitions, as is done at l.\ref{l:log-trans-closure} of
Algorithm~\ref{alg:partref-trans}.

The correctness of this algorithm relies on the fact that equality on states is
invariant under the partial closure that is added. Indeed, we can see that the
DFA $A'$ obtained in Algorithm~\ref{alg:partref-trans} is language equivalent
to the input DFA $A$ if we consider the alphabet letters added as words. If
$\delta^T(q,a_T) = q'$ for some $a_T \in \Sigma^T$, then $a_T= a^{2^j}$ for some
$a\in \Sigma$ and $j \in [0, \lfloor \log n\rfloor ]$. 
By
construction there is a sequence $q_0, \dots q_{k}$ such that $q_0 = q$,
$q_{i+1} = \delta(q_i,a)$ and $q_{k} = q'$. 

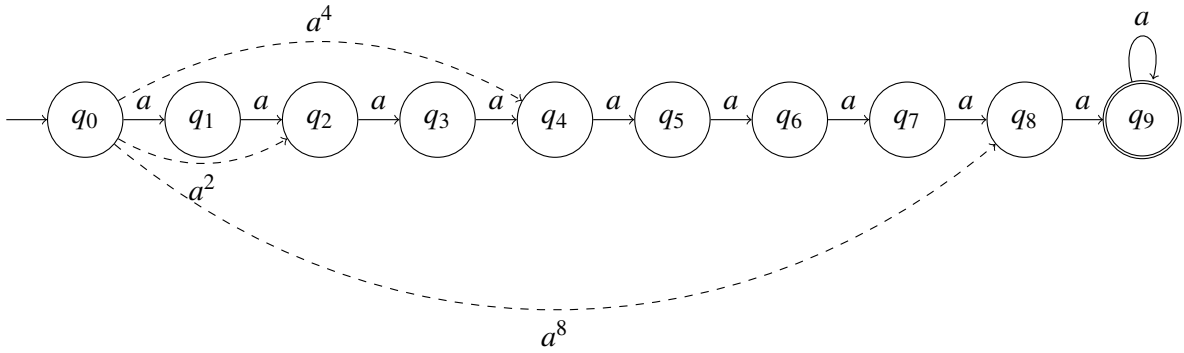
\begin{figure}
  \resizebox{\textwidth}{!}{
\begin{tikzpicture}[node distance= 1.5cm, initial text = ]
  \node[state,initial] (0) at (0,0) {$q_0$};
  \node[state] (1) [right of=0]{$q_1$};
  \node[state] (2) [right of=1]{$q_2$};
  \node[state] (3) [right of=2]{$q_3$};
  \node[state] (4) [right of=3]{$q_4$};
  \node[state] (5) [right of=4]{$q_5$};
  \node[state] (6) [right of=5]{$q_6$};
  \node[state] (7) [right of=6]{$q_7$};
  \node[state] (8) [right of =7] {$q_8$};
  \node[state, accepting] (9) [right of =8] {$q_9$};

  \path[->]
  (0) edge node [above]{$a$} (1) 
  (1) edge node [above]{$a$} (2) 
  (2) edge node [above]{$a$} (3) 
  (3) edge node [above]{$a$} (4) 
  (4) edge node [above]{$a$} (5) 
  (5) edge node [above]{$a$} (6) 
  (6) edge node [above]{$a$} (7) 
  (7) edge node [above]{$a$} (8) 
  (8) edge node [above]{$a$} (9) 
  (9) edge[loop above] node [above]{$a$} (9);

  \path[->, dashed] 
  (0) edge [bend right=30] node[below] {$a^2$} (2) (0) edge [bend left]
  node[above] {$a^4$} (4) (0) edge [bend right=40] node[below] {$a^8$} (8);
  \end{tikzpicture}} \caption{The DFA $A=(\{q_0, \dots , q_9\}, \{a\}, \delta,
  \{q_9\}, q_0)$ with the extra partial transitive closure from $q_0$ added in dashed
  arrows.\label{fig:dfalong}}
\end{figure}

This approach helps in the case of long paths with the same letter. Consider the
DFA $A$ from Figure~\ref{fig:dfalong}. This DFA accepts all words $a^j$ with
$j > 8$. Any parallel partition refinement algorithm would need at least $8$
iterations to conclude that $q_0$ is not the same as $q_1$.  However, building
the partial transitive closure only requires a logarithmic number of parallel
iterations. With this partial transitive closure added, a partition refinement
algorithm can in the first iteration conclude that $q_0$ is different from $q_1$
since the transition with $a^8$ leads to different states.  

\section{Experiments}
\label{sec:results}

In this section, we discuss the results of our implementations. We benchmarked
the implementations with respect to three families of DFAs: Fibonacci DFAs
from~\cite{castiglione2008hopcroft}, bit-splitters $\mathcal{B}_k$ derived
from~\cite{groote2023lowerbounds}, and DFAs derived from a subset of the VLTS
benchmark set.

\subsection{Benchmarks}

\paragraph{Fibonacci DFAs:} The first family of DFAs we use for benchmarking
consists of so-called Fibonacci automata. These are simple automata with only a
unary alphabet. However, they exhibit very particular behaviour. As witnessed
in~\cite{castiglione2008hopcroft}, these automata are notoriously hard for
partition refinement and the number of iterations of any partition refinement
algorithm is $n$. The automata are called Fibonacci automata due to the close
correspondence with Fibonacci words over the binary alphabet, which are defined
inductively as follows: the base cases are $w_0 = 1$, and $w_1 = 0$, and for every $i\in \Nat$,
$w_{n+1} = w_nw_{n-1}$. This gives the following sequence:
\begin{align*}
  w_2 &= 01 \\
  w_3 &= 010 \\ 
  w_4 &= 01001 \\
  w_5 &= 01001010 \\
  \ldots
\end{align*}

For every $n\in \Nat$, we define the automaton $\Fib_n = (Q, \{a\}, \delta, q_0, F)$ as follows,
with $w_n[i]$ referring to the $i$-th bit in the bit sequence $w_n$:
\begin{itemize}
  \item the set of states is $Q =\{q_i \mid i \in [0,|w_n| ] \}$; 
\item the transition function is $\delta(q_i,a) = q_{i+1\mod |w_n|}$;
  
  \item the set of final states is $F = \{q_i \mid q_i \in Q\text{ and } w_n[i] = 1\}$.
\end{itemize}
\newcommand{\bit}[1]{\mathtt{#1}}

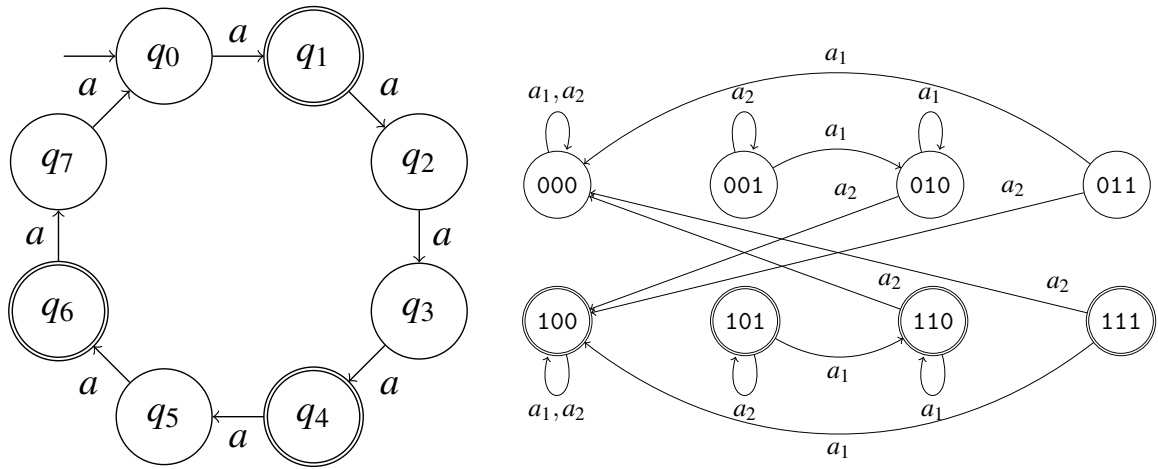
\begin{figure}
  \begin{center}
    \resizebox{0.39\textwidth}{!}{
  \begin{tikzpicture}[node distance= 1.5cm, initial text = ]
    \node[state,initial] (0) at (0,0) {$q_0$};
    \node[state, accepting] (1) [right of=0]{$q_1$};
    \node[state] (2) [below right of=1]{$q_2$};
    \node[state] (3) [below of=2]{$q_3$};
    \node[state,,accepting] (4) [below left of=3]{$q_4$};
    \node[state] (5) [left of=4]{$q_5$};
    \node[state,accepting] (6) [above left of=5]{$q_6$};
    \node[state] (7) [above of=6]{$q_7$};

    \path[->]
    (0) edge node [above]{$a$} (1) 
    (1) edge node [above right ]{$a$} (2)
    (2) edge node [right]{$a$} (3)
    (3) edge node [below right]{$a$} (4) 
    (4) edge node [below]{$a$} (5)
    (5) edge node [below left]{$a$} (6)
    (6) edge node [left]{$a$} (7)
    (7) edge node [above left]{$a$} (0);
    \end{tikzpicture}}
\resizebox{0.6\textwidth}{!}{
    \begin{tikzpicture}[node distance=1.75cm] 

    
      \node[state] (000) {$\bit{000}$};
    
      \node[state, right=of 000]  (001) {$\bit{001}$};
      \node[state, right= of 001] (010) {$\bit{010}$};
      \node[state, right= of 010] (011) {$\bit{011}$};
      \node[state, accepting, below= 1cm of 000]  (100) {$\bit{100}$};
      \node[state, accepting, right=of 100]  (101) {$\bit{101}$};
      \node[state, accepting, right=of 101]  (110) {$\bit{110}$};
      \node[state, accepting, right=of 110]  (111) {$\bit{111}$};
    
        
      \node[below left =0.5cm and 0.5cm of 000]  (lineleft) {};
      \node[above right=0.5cm and 0.5cm of 111] (lineright) {};
        
      \path[->] 
      (000) edge [loop above] node {$a_1, a_2$} (000)
      
      (001) edge [loop above] node {$a_2$} (001)
      (001) edge [bend left]  node [above] {$a_1$} (010)
      
      (010) edge node [pos=0.1, above left] {$a_2$} (100)
      (010) edge [loop above] node [above] {$a_1$} (010)
      
      (011) edge node [pos=0.1, above left]{$a_2$} (100)
      (011) edge [bend right=38] node[above] {$a_1$} (000)
      
      (100) edge [loop below] node{$a_1, a_2$} (100)
      
      (101) edge [bend right] node[below] {$a_1$} (110)
      (101) edge [loop below] node[below] {$a_2$} (101)
      
      (110) edge node[pos=0.1, above right] {$a_2$} (000)
      (110) edge [loop below] node [below] {$a_1$} (110)
      
      (111) edge node[pos=0.1, above right] {$a_2$} (000)
      (111) edge[bend left=38] node[below] {$a_1$} (100)
       ;
    \end{tikzpicture}}
    \end{center}
    \caption{The DFA $\Fib_5$ on the left, and the DFA $\mathcal{B}_3$ on the
    right.\label{fig:dfas}} 
\end{figure}

\vspace{-8pt}

\paragraph{Bit-splitters:} 
The second family of automata consists of the so-called \emph{bit-splitters}
$\mathcal{B}_n$. For $n\in \Nat$, the bit-splitter $\mathcal{B}_n$ is a
deterministic automaton with $2^n$ states and an alphabet consisting of $n{-}1$
symbols. By construction, during partition refinement, every time a block can be
split, it is split in two blocks of equal size. This property makes the family
inherently hard for partition refinement algorithms. However, the parallel
algorithm requires only a logarithmic number of iterations to compute the
minimal DFA. The bit splitter $\mathcal{B}_3$ is given in Figure~\ref{fig:dfas}.

The family does not contain an initial state, and comes from the setting of
Labelled Transition Systems (LTSs).
An LTS is a graph structure with a finite number of states and transitions
between states, with each transition having an action label.

Since it is a hard example for partition refinement, we use it for this
purpose and allow the absence of an initial state.
In the following, states $\sigma$ represent bit sequences of length $n$,
i.e., $\sigma \in \{0, 1\}^n$.
We define $\mathcal{B}_1= (Q_1,
\Sigma_1, \delta_1, F_1)$, where $Q_1 = \{\bit0, \bit1\}$, $\Sigma_1 =
\emptyset$ and $F_1=\{\bit1\}$. Given the automaton $\mathcal{B}_n = (Q_n, \Sigma_n,
\delta_n, F_n)$ for some $n \in \Nat$,
we define $\mathcal{B}_{n+1}=(Q_{n+1},
\Sigma_{n+1}, \delta_{n+1}, F_{n+1})$, such that:
\begin{itemize}
  \item The set of states contains two copies of $Q_n$, i.e., $Q_{n+1} = \{
  \bit0\sigma, \bit1\sigma \mid \sigma\in Q_n \}$,
  \item One fresh symbol $a_n \not\in \Sigma_n$ is added to the alphabet: $\Sigma_{n+1} = \Sigma_{n}
  \cup \{ a_n \}$,
  \item The transition function $\delta_{n+1}$ is defined such that for each
  $a_m \in \Sigma_n$, and state $\bit{b}\sigma \in Q_{n+1}$, it maintains the
  behaviour of $\mathcal{B}_{n}$, i.e., $\delta_{n+1}(a_m, \bit{b}\sigma) =
  \bit{b}\delta_n(a_m, \sigma)$. For the fresh symbol $a_n \in \Sigma_{n+1}
  \setminus \Sigma_n$, $\delta_{n{+}1}$ is extended as follows, with $\bit{\bar{b}}$ being
  the bit flipped version of $\bit{b}$, i.e., $\bit{b} = 0 \iff \bit{\bar{b}} = 1$:
  \[
    \delta_{n{+}1}(\bit{b}\sigma, a_n)= \begin{cases}
      \bit{\bar{b}} \bit0^n & \text{If } \sigma[0] = 1,\\
      \bit{b}\sigma & \text{otherwise.}
    \end{cases}
  \]
  \item the set of accepting states is $F_{n+1} = \{\bit1 \sigma \mid \sigma \in
  Q_n \}$.
\end{itemize}

As previously mentioned, bit-splitter DFAs are constructed in such a way that they are inherently hard to
minimize by partition refinement. Each bit-splitter $\mathcal{B}_{n+1}$ combines
two copies of the bit-splitter $\mathcal{B}_n$, with the transition
function defined in such a way that each possible split divides an existing block in two
blocks of the same size. This results in a DFA in which the amount of required work for splitting is large,
since each split involves moving many states to the new block. However, because in each split a block is split in two parts of the same size, the number of sequential splits
needed is smaller than for the Fibonacci automata.

\vspace{-8pt}

\paragraph{VLTSs:}
Lastly, we benchmark our implementations against the VLTS benchmark
suite.\footnote{\url{https://cadp.inria.fr/resources/vlts} (visited on: 04-2024).} The VLTS acronym stands
for \emph{Very Large Transition Systems}. This suite consists of LTSs that
originate from modelling protocols and concurrent systems. Some of the
benchmarks are from case studies from industrial systems. 

The transition
relation of an LTS does not need to be deterministic, nor complete. We turn an LTS into a
DFA by first making the LTS deterministic such that each state has at most one
outgoing transition for every label, using the powerset construction
algorithm~\cite{powerset}. To convert the deterministic LTS to a DFA, we need to
complete the transition function and define which states are accepting. We
define all the states as accepting and add one new non-accepting state $\bot$.
For each state $q$ and label $a$ for which there exists no transition with that
label from $q$, we add a new transition labelled $a$ to $\bot$, i.e.
$\delta(q,a) = \bot$. This completes the transition function and creates a DFA
accepting all the words corresponding with a path through the original LTS. 

Due to state-space explosion we were not able to make all VLTS benchmarks
deterministic. We used all benchmarks for which the computation to make them
deterministic took less than ten minutes.

\subsection{Results}
\label{sec:results}
The algorithms were implemented in CUDA \CC and compiled using the CUDA toolkit 12.2, with
the implementation of \texttt{sortPR} using the Thrust library for sorting and
computing the adjacent differences and inclusive scans~\cite{thrust}.
Experiments were conducted on a device running Linux Mint 20, equipped with an
NVIDIA TITAN RTX GPU with 24 GB of memory and 4,608 cores. Such a GPU can manage
trillions of light-weight threads. Thanks to fast context switching between threads, a GPU can
 typically handle a few hundred thousand threads as if they execute in parallel. 

The reported times are the average of five separate runs. Benchmarks that did
not finish within five minutes were aborted, in which case we registered a
timeout `t/o'. Benchmarks for which there was not enough memory are indicated by
`OoM'.

\vspace{-8pt}

\paragraph{Transitive approach:}
The results of running Algorithm~\ref{alg:fulltrans} on the Fibonacci automata
are given in Table~\ref{table:trans}. As expected, the number of threads used to
compute the transitive closure in parallel grows very quickly. Although the
number of iterations of the algorithm is indeed logarithmic, the available
parallelism is not sufficient to lead to logarithmic run times. We only use
this set of small Fibonacci automata for this algorithm. It already suggests
that for a relatively small amount of states (~$\sim 100$), obtaining the required resources is
already infeasible. The other benchmarks are almost completely out of range of
the algorithm.

\input{results-table-transitive.tex}

\vspace{-8pt}

\paragraph{Partition refinement algorithms:} The results of running the parallel
partition refinement algorithms,
$\texttt{naivePR}$~(Algorithm~\ref{alg:partref}),
$\texttt{sortPR}$~(Algorithm~\ref{alg:sort}), and
$\texttt{transPR}$~(Algorithm~\ref{alg:partref-trans}) are given in
Table~\ref{table:concur} and Table~\ref{table:vlts} for the different
benchmarks.

First, we observe in Table~\ref{table:concur} that on the Fibonacci automata the
$\texttt{naivePR}$ performs better than $\texttt{sortPR}$. This can be explained
by the fact that the number of iterations is $n$ for both algorithms, while each
iteration in $\texttt{sortPR}$ is slower than in $\texttt{naivePR}$. Another
interesting observation here is that for all benchmarks $\Fib_{18}, \dots,
\Fib_{28}$ the run time of the algorithm $\texttt{naivePR}$ scales linearly with
the number of states $n$. Since the number of parallel iterations is $n{-}2$ for
all these benchmarks, each parallel iteration processing up to $\sim 500k$
states took a similar amount of time. In other words, the GPU was able to run
around $500k$ threads as if they ran in parallel. This confirms the statement
about fast context switching at the beginning of Section~\ref{sec:results}.

Finally, for the Fibonacci automata, we see that $\texttt{transPR}$ performs
significantly better on this benchmark. This can be explained by the fact that the
partial transitive closure reduces the number of iterations of the algorithm
significantly. 

The results on the bit-splitter automata in Table~\ref{table:concur} show that
the improvement of $\texttt{transPR}$ does not work on all automata. The high
number of alphabet letters together with the structure of the automata make the
transitive closure less effective, making $\texttt{naivePR}$ much faster.

For the VLTS benchmark set we see the power of $\texttt{sortPR}$ in Table~\ref{table:vlts}. In some
benchmarks, like `vasy\_69\_520' the algorithm performs significantly better. In
these examples, it helps that in $\texttt{sortPR}$, in each iteration a block
can be split into many subblocks, which is not the case in the other algorithms.

Since the VLTS benchmarks originate from communication
protocols and concurrent systems, the success of \texttt{sortPR} suggests that
for DFAs that represent `real' systems, this algorithm is a solid choice for
efficient DFA minimization. However, the experiments with the Fibonacci and
bit-splitter families of DFAs demonstrate room for improvement.

\input{results-table-concur.tex}
\input{results-table-vltss.tex}

\section{Conclusions \& Future work}
We implemented and compared different parallel algorithms for DFA minimization
on GPUs. We find that the $NC$ algorithm~$\texttt{trans}$ with parallel
logarithmic run-time does not scale well because of the large number of
resources needed. Instead, we find that the partition refinement algorithms
perform better. This might be seen as contradictory since these partition
refinement algorithms have inherently linear parallel run-times. 

When comparing the different partition refinement algorithms, the structure of the
input DFA is of high influence. The trade-off is that in $\texttt{sortPR}$ each
iteration takes more time than in $\texttt{naivePR}$, but in \texttt{sortPR},
an iteration has the potential to lead to a block
being split into more than two subblocks. When this happens sufficiently often,
fewer iterations are needed. This leads to
$\texttt{sortPR}$ being slower in cases where the number of iterations is high.
In other benchmarks, it leads to fewer iterations and thereby a significant
speed-up. 

Finally, we showed a way to incorporate a partial transitive closure in
partition refinement algorithms. We showed that for a specific class of DFAs
this approach leads to logarithmic run-times, where every partition refinement
algorithm is inherently linear. 

As future work it would be interesting to further investigate sublinear time
parallel algorithms for DFA minimization. Specifically, there are two key
questions that come to mind. The first question is: what is a reasonable number
of parallel processes necessary for a poly-logarithmic time parallel algorithm?
It seems feasible to use a similar argument as in~\cite{khuller1994parallel} to
get a superlinear lower bound. However, the gap between the $O(n^{2\omega})$
processors%
\footnote{If matrix multiplication can be computed in time $O(n^\omega)$,
currently best known bounds $\omega \leq 2.372\dots$.} used in
Algorithm~\ref{alg:fulltrans} remains large. The second question is: can a method
such as the one presented in this paper using the partial transitive closure be
implemented in such a way that the run
time will be sublinear with high probability, i.e. any parallel run-time
$O(n^{1{-}\epsilon})$ for some $\epsilon \geq 0$. A good starting point
would be the recent work on parallel reachability
algorithms~\cite{ullman1990high,fimeman2018nearwork,jambulapati2019parallel}.

\newpage
\bibliographystyle{eptcs}
\bibliography{references}
\end{document}

%% file: results-table-transitive.tex
\begin{table}[tb]
  \centering
  \begin{tabular}{lc|rr|rr}
    Name & $N$ & Iterations & Time (ms) & Memory(Mb) & \#threads \\ \hline
    $\Fib_4$ & 8 & 3 & 0.3 & 0 & 589,824\\
    $\Fib_5$ & 13 & 4 & 0.7 & 0 & 6,230,016\\
    $\Fib_6$ & 21 & 5 & 7.8 & 0 & 88,510,464\\
    $\Fib_7$ & 34 & 5 & 159.9 & 0 & 1,620,545,536\\
    $\Fib_8$ & 55 & 6 & 3,034.9 & 10 & 27,955,840,000\\
    $\Fib_9$  & 89 & 7 & 66,846.7 & 60 & 498,865,340,416\\
    $\Fib_{10}$ & 144 & t/o & t/o & 412 & 8,943,640,510,464
  \end{tabular}
  \caption{Results of running the algorithm $\texttt{trans}$ on the Fibonacci automata.\label{table:trans}}
  \end{table}

%% file: results-table-concur.tex
  \begin{table}[tb]
    \centering
    \resizebox{\textwidth}{!}{
  \begin{tabular}{l|rrr|rrr|rrr}
    & \multicolumn{3}{c|}{\textbf{Benchmark metrics}} &
      \multicolumn{3}{c|}{\textbf{Times (ms)}} &
      \multicolumn{3}{c}{\textbf{Iterations}}\\
    Name & \multicolumn{1}{c}{$N$} & \multicolumn{1}{c}{$|\Sigma|$} &
           \multicolumn{1}{c|}{Size output} & \multicolumn{1}{c}{$\mathtt{naivePR}$} & 
           \multicolumn{1}{c}{$\mathtt{sortPR}$} & \multicolumn{1}{c|}{$\mathtt{transPR}$} &
           \multicolumn{1}{c}{$\mathtt{naivePR}$} & \multicolumn{1}{c}{$\mathtt{sortPR}$} &
           \multicolumn{1}{c}{$\mathtt{transPR}$}
  \\ \hline
  $\Fib_{20}$        & 17,711     & 1  & 17,711                         & 308.8                       & 3,909.2   & 1.7     & 17,710   & 17,710                   & 14                      \\
  $\Fib_{21}$        & 28,657     & 1  & 28,657                         & 494.2                       & 6,374.2   & 2.4     & 28,656   & 28,656                   & 25                      \\
  $\Fib_{22}$        & 46,368     & 1  & 46,368                         & 778.7                       & 11,712.1  & 4.1     & 46,367   & 46,367                   & 61                      \\
  $\Fib_{23}$        & 75,025     & 1  & 75,025                         & 1,241.3                     & 21,366.6  & 8.0     & 75,024   & 75,024                   & 101                     \\
  $\Fib_{24}$        & 121,393    & 1  & 121,393                        & 2,006.7                     & 34,793.1  & 12.5    & 121,392  & 121,392                  & 104                     \\
  $\Fib_{25}$        & 196,418    & 1  & 196,418                        & 3,251.3                     & 64,411.7  & 18.3    & 196,417  & 196,417                  & 138                     \\
  $\Fib_{26}$        & 317,811    & 1  & 317,811                        & 5,277.8                     & 178,367.4 & 49.8    & 317,810  & 317,810                  & 102                     \\
  $\Fib_{27}$        & 514,229    & 1  & 514,229                        & 8,607.7                     & t/o       & 96.1    & 514,228  & \multicolumn{1}{r}{t/o} & 268                     \\
  $\Fib_{28}$        & 832,040    & 1  & 832,040                        & 22,723.0                    & t/o       & 178.4   & 832,039  & \multicolumn{1}{r}{t/o} & 299                     \\
  $\Fib_{29}$        & 1,346,269  & 1  & 1,346,269                      & 59,510.8                    & t/o       & 726.9   & 1,346,268 & \multicolumn{1}{r}{t/o} & 755                     \\
  $\Fib_{30}$        & 2,178,309  & 1  & 2,178,309                      & 141,601.0                   & t/o       & 1,109.3 & 2,178,308 & \multicolumn{1}{r}{t/o} & 914                     \\
  \hline
  $\mathcal{B}_{15}$ & 32,768     & 14 & 32,768                         & 0.8                         & 25.8      & 1.7     & 14      & 14                      & 2                       \\
  $\mathcal{B}_{16}$ & 65,536     & 15 & 65,536                         & 1.4                         & 29.7      & 3.7     & 15      & 15                      & 2                       \\
  $\mathcal{B}_{17}$ & 131,072    & 16 & 131,072                        & 2.6                         & 54.3      & 9.4     & 16      & 16                      & 2                       \\
  $\mathcal{B}_{18}$ & 262,144    & 17 & 262,144                        & 5.0                         & 107.2     & 25.6    & 17      & 17                      & 2                       \\
  $\mathcal{B}_{19}$ & 524,288    & 18 & 524,288                        & 9.6                         & 235.7     & 60.9    & 18      & 18                      & 2                       \\
  $\mathcal{B}_{20}$ & 1,048,576  & 19 & 1,048,576                      & 19.3                        & 520.2     & 139.8   & 19      & 19                      & 2                       \\
  $\mathcal{B}_{21}$ & 2,097,152  & 20 & 2,097,152                      & 39.8                        & 1,148.6   & 312.2   & 20      & 20                      & 2                       \\
  $\mathcal{B}_{22}$ & 4,194,304  & 21 & 4,194,304                      & 82.6                        & 2,538.5   & 728.7   & 21      & 21                      & 2                       \\
  $\mathcal{B}_{23}$ & 8,388,608  & 22 & 8,388,608                      & 170.3                       & 5,612.7   & 1,612.1 & 22      & 22                      & 2                       \\
  $\mathcal{B}_{24}$ & 16,777,216 & 23 & 16,777,216                     & 352.6                       & 12,351.8  & OoM     & 23      & 23                      & OoM                     \\
  $\mathcal{B}_{25}$ & 33,554,432 & 24 & 33,554,432                     & 737.4                       & 27,092.2  & OoM     & 24      & 24                      & OoM \\
  $\mathcal{B}_{26}$ & 67,108,864 & 25 & 67,108,864                     & 1,541.5                     & 59,203.8  & OoM     & 25      & 25                      & OoM
  \end{tabular}}
  \caption{Results of running the partition refinement algorithms on the $\Fib$ and $\mathcal{B}$ benchmark set.\label{table:concur}}
  \end{table}

%% file: results-table-vltss.tex

\begin{table}[tb]
  \centering
  \resizebox{\textwidth}{!}{
  \begin{tabular}{l|rrr|rrr|rrr}
    & \multicolumn{3}{c|}{\textbf{Benchmark metrics}} & \multicolumn{3}{c}{\textbf{Times (ms)}} &\multicolumn{3}{|c}{\textbf{Iterations}}\\
    Name & 
    \multicolumn{1}{c}{$N$} & \multicolumn{1}{c}{$|\Sigma|$} & \multicolumn{1}{c|}{Size output} &
    \multicolumn{1}{c}{$\mathtt{naivePR}$} &\multicolumn{1}{c}{$\mathtt{sortPR}$} & \multicolumn{1}{c|}{$\mathtt{transPR}$} & 
    \multicolumn{1}{c}{$\mathtt{naivePR}$} & \multicolumn{1}{c}{$\mathtt{sortPR}$} & \multicolumn{1}{c}{$\mathtt{transPR}$}
    \\ \hline
    cwi\_1\_2         & 4,448     & 26     & 2,416   & 5.4       & 66.7     & 25.1      & 308    & 38    & 621   \\
    cwi\_2416\_17605  & 503       & 15     & 58      & 0.8       & 38.2     & 0.4       & 40     & 40    & 8     \\
    cwi\_3\_14        & 63        & 2      & 63      & 1.2       & 9.1      & 0.4       & 61     & 61    & 8     \\
    vasy\_0\_1        & 92        & 2      & 10      & 0.2       & 3.9      & 0.4       & 6      & 5     & 5     \\
    vasy\_1\_4        & 6,087     & 6      & 29      & 0.4       & 8.5      & 0.9       & 15     & 7     & 20    \\
    vasy\_10\_56      & 10,850    & 12     & 2113    & 8.7       & 40.2     & 30.9      & 519    & 33    & 791   \\
    vasy\_1112\_5290  & 1,112,491 & 23     & 266     & 135.4     & 386.8    & 2,049.2   & 246    & 4     & 231   \\
    vasy\_157\_297    & 157,605   & 235    & 4,290   & 455.1     & 1,736.3  & 11,312.0  & 1,049   & 27    & 1,306  \\
    vasy\_164\_1619   & 109,911   & 37     & 1,025   & 69.9      & 50.5     & 823.4     & 770    & 4     & 766   \\
    vasy\_166\_651    & 393,147   & 211    & 392,175 & 159,265.6 & 1,070.6  & t/o       & 175,764 & 19    & t/o   \\
    vasy\_18\_73      & 419,664   & 17     & 31,952  & 1,586.1   & 305.2    & 34,055.2  & 13,343  & 27    & 18,444 \\
    vasy\_25\_25      & 25,218    & 25,216 & 25,218  & 262,878.6 & 3,502.7  & t/o       & 25,217  & 2     & t/o   \\
    vasy\_386\_1171   & 355,790   & 73     & 114     & 36.9      & 489.4    & 766.0     & 58     & 8     & 113   \\
    vasy\_40\_60      & 40,007    & 3      & 40,007  & 331.6     & 8,391.5  & 845.2     & 20,004  & 20002 & 20,004 \\
    vasy\_5\_9        & 5,088     & 31     & 138     & 2.2       & 14.3     & 7.0       & 113    & 5     & 124   \\
    vasy\_574\_13561  & 574,058   & 141    & 3,578   & 2,332.2   & 976.5    & 64,312.6  & 2,351   & 5     & 2,634  \\
    vasy\_6120\_11031 & 3,190,785 & 125    & 5,216   & 13,186.6  & 21,886.0 & t/o       & 2,373   & 21    & t/o   \\
    vasy\_65\_2621    & 65,538    & 72     & 65,537  & 2,591.8   & 38.3     & 47,568.0  & 36,575  & 4     & 38,999 \\
    vasy\_66\_1302    & 209,791   & 81     & 208,419 & 42,864.9  & 96.0     & t/o       & 179,861 & 8     & t/o   \\
    vasy\_69\_520     & 74,958    & 135    & 74,958  & 7,223.0   & 124.2    & 181,611.4 & 49,723  & 12    & 74,667 \\
    vasy\_720\_390    & 87,741    & 49     & 3,279   & 176.0     & 57.1     & 2,961.7   & 2,936   & 5     & 2,950  \\
    vasy\_8\_24       & 20,306    & 11     & 560     & 5.9       & 26.8     & 22.1      & 282    & 17    & 348   \\
    vasy\_8\_38       & 8,922     & 81     & 220     & 5.7       & 44.1     & 31.5      & 174    & 5     & 215   \\
    vasy\_83\_325     & 393,147   & 211    & 392,175 & 162,495.0 & 1,074.4  & t/o     & 173,218 & 19    & t/o  
    \end{tabular}}
  \caption{Results of running the partition refinement algorithms on the VLTS benchmark set.\label{table:vlts}}
  \end{table}

%% file: main.bbl
\begin{thebibliography}{10}
\providecommand{\bibitemdeclare}[2]{}
\providecommand{\surnamestart}{}
\providecommand{\surnameend}{}
\providecommand{\urlprefix}{Available at }
\providecommand{\url}[1]{\texttt{#1}}
\providecommand{\href}[2]{\texttt{#2}}
\providecommand{\urlalt}[2]{\href{#1}{#2}}
\providecommand{\doi}[1]{doi:\urlalt{https://doi.org/#1}{#1}}
\providecommand{\eprint}[1]{arXiv:\urlalt{https://arxiv.org/abs/#1}{#1}}
\providecommand{\bibinfo}[2]{#2}

\bibitemdeclare{article}{balcazar1992}
\bibitem{balcazar1992}
\bibinfo{author}{J.~\surnamestart Balc{\'a}zar\surnameend},
  \bibinfo{author}{J.~\surnamestart Gabarro\surnameend} \&
  \bibinfo{author}{M.~\surnamestart Santha\surnameend} (\bibinfo{year}{1992}):
  \emph{\bibinfo{title}{Deciding bisimilarity is {P}-complete}}.
\newblock {\slshape \bibinfo{journal}{Formal aspects of computing}}
  \bibinfo{volume}{4}(\bibinfo{number}{1}), pp. \bibinfo{pages}{638--648},
  \doi{10.1007/BF03180566}.

\bibitemdeclare{incollection}{thrust}
\bibitem{thrust}
\bibinfo{author}{N.~\surnamestart Bell\surnameend} \&
  \bibinfo{author}{J.~\surnamestart Hoberock\surnameend}
  (\bibinfo{year}{2012}): \emph{\bibinfo{title}{{Thrust: A
  Productivity-Oriented Library for CUDA}}}.
\newblock In: {\slshape \bibinfo{booktitle}{{GPU Computing Gems Jade
  Edition}}}, chapter~\bibinfo{chapter}{26}, \bibinfo{publisher}{Morgan
  Kaufmann Publishers Inc.}, pp. \bibinfo{pages}{359--371},
  \doi{10.1016/C2010-0-68654-8}.

\bibitemdeclare{inproceedings}{castiglione2008hopcroft}
\bibitem{castiglione2008hopcroft}
\bibinfo{author}{G.~\surnamestart Castiglione\surnameend},
  \bibinfo{author}{A.~\surnamestart Restivo\surnameend} \&
  \bibinfo{author}{M.~\surnamestart Sciortino\surnameend}
  (\bibinfo{year}{2008}): \emph{\bibinfo{title}{Hopcroft's Algorithm and Cyclic
  Automata}}.
\newblock In \bibinfo{editor}{C.~\surnamestart Mart{\'{\i}}n-Vide\surnameend},
  \bibinfo{editor}{F.~\surnamestart Otto\surnameend} \&
  \bibinfo{editor}{H.~\surnamestart Fernau\surnameend}, editors: {\slshape
  \bibinfo{booktitle}{Proc.\ of LATA 2008}}, {\slshape \bibinfo{series}{LNCS}}
  \bibinfo{volume}{5196}, \bibinfo{publisher}{Springer}, pp.
  \bibinfo{pages}{172--183}, \doi{10.1007/978-3-540-88282-4_17}.

\bibitemdeclare{article}{cho1992parallelcoarsestset}
\bibitem{cho1992parallelcoarsestset}
\bibinfo{author}{S.~\surnamestart Cho\surnameend} \& \bibinfo{author}{D.T.
  \surnamestart Huynh\surnameend} (\bibinfo{year}{1992}):
  \emph{\bibinfo{title}{The parallel complexity of coarsest set partition
  problems}}.
\newblock {\slshape \bibinfo{journal}{Information Processing Letters}}
  \bibinfo{volume}{42}(\bibinfo{number}{2}), pp. \bibinfo{pages}{89--94},
  \doi{10.1016/0020-0190(92)90095-D}.

\bibitemdeclare{inproceedings}{fimeman2018nearwork}
\bibitem{fimeman2018nearwork}
\bibinfo{author}{J.T. \surnamestart Fineman\surnameend} (\bibinfo{year}{2018}):
  \emph{\bibinfo{title}{Nearly Work-Efficient Parallel Algorithm for Digraph
  Reachability}}.
\newblock In: {\slshape \bibinfo{booktitle}{Proc.\ of STOC 2018}},
  \bibinfo{publisher}{ACM}, p. \bibinfo{pages}{457–470},
  \doi{10.1145/3188745.3188926}.

\bibitemdeclare{article}{groote2023lowerbounds}
\bibitem{groote2023lowerbounds}
\bibinfo{author}{J.F. \surnamestart Groote\surnameend}, \bibinfo{author}{J.J.M.
  \surnamestart Martens\surnameend} \& \bibinfo{author}{E.P. \surnamestart
  de~Vink\surnameend} (\bibinfo{year}{2023}):
  \emph{\bibinfo{title}{{Lowerbounds for bisimulation by partition
  refinement}}}.
\newblock {\slshape \bibinfo{journal}{{Logical Methods in Computer Science}}}
  \bibinfo{volume}{{Volume 19, Issue 2}}, \doi{10.46298/lmcs-19(2:10)2023}.

\bibitemdeclare{article}{Hillis86}
\bibitem{Hillis86}
\bibinfo{author}{W.D. \surnamestart Hillis\surnameend} \&
  \bibinfo{author}{G.L.~Steele \surnamestart Jr.\surnameend}
  (\bibinfo{year}{1986}): \emph{\bibinfo{title}{Data Parallel Algorithms}}.
\newblock {\slshape \bibinfo{journal}{Communications of the {ACM}}}
  \bibinfo{volume}{29}(\bibinfo{number}{12}), pp. \bibinfo{pages}{1170--1183},
  \doi{10.1145/7902.7903}.

\bibitemdeclare{incollection}{hopcroft1971DFAmin}
\bibitem{hopcroft1971DFAmin}
\bibinfo{author}{J.~\surnamestart Hopcroft\surnameend} (\bibinfo{year}{1971}):
  \emph{\bibinfo{title}{An $n \log n$ algorithm for minimizing states in a
  finite automaton}}.
\newblock In \bibinfo{editor}{Z.~\surnamestart Kohavi\surnameend} \&
  \bibinfo{editor}{A.~\surnamestart Paz\surnameend}, editors: {\slshape
  \bibinfo{booktitle}{Theory of Machines and Computations}},
  \bibinfo{publisher}{Academic Press}, pp. \bibinfo{pages}{189--196},
  \doi{10.1016/b978-0-12-417750-5.50022-1}.

\bibitemdeclare{book}{jaja1992introduction}
\bibitem{jaja1992introduction}
\bibinfo{author}{J.~\surnamestart J\'{a}J\'{a}\surnameend}
  (\bibinfo{year}{1992}): \emph{\bibinfo{title}{An introduction to parallel
  algorithms}}.
\newblock \bibinfo{publisher}{Addison Wesley Longman Publishing Co., Inc.},
  \bibinfo{address}{USA}.

\bibitemdeclare{inproceedings}{jambulapati2019parallel}
\bibitem{jambulapati2019parallel}
\bibinfo{author}{A.~\surnamestart Jambulapati\surnameend},
  \bibinfo{author}{Y.P. \surnamestart Liu\surnameend} \&
  \bibinfo{author}{A.~\surnamestart Sidford\surnameend} (\bibinfo{year}{2019}):
  \emph{\bibinfo{title}{Parallel reachability in almost linear work and square
  root depth}}.
\newblock In: {\slshape \bibinfo{booktitle}{Proc.\ of FOCS 2019}},
  \bibinfo{organization}{IEEE}, pp. \bibinfo{pages}{1664--1686},
  \doi{10.1109/FOCS.2019.00098}.

\bibitemdeclare{article}{khuller1994parallel}
\bibitem{khuller1994parallel}
\bibinfo{author}{S.~\surnamestart Khuller\surnameend} \&
  \bibinfo{author}{U.~\surnamestart Vishkin\surnameend} (\bibinfo{year}{1994}):
  \emph{\bibinfo{title}{On the parallel complexity of digraph reachability}}.
\newblock {\slshape \bibinfo{journal}{Information Processing Letters}}
  \bibinfo{volume}{52}(\bibinfo{number}{5}), pp. \bibinfo{pages}{239--241},
  \doi{10.1016/0020-0190(94)00153-7}.

\bibitemdeclare{article}{martens2022linear}
\bibitem{martens2022linear}
\bibinfo{author}{J.J.M. \surnamestart Martens\surnameend},
  \bibinfo{author}{J.F. \surnamestart Groote\surnameend}, \bibinfo{author}{L.B.
  \surnamestart Haak\surnameend}, \bibinfo{author}{P.~\surnamestart
  Hijma\surnameend} \& \bibinfo{author}{A.J. \surnamestart Wijs\surnameend}
  (\bibinfo{year}{2022}): \emph{\bibinfo{title}{Linear parallel algorithms to
  compute strong and branching bisimilarity}}.
\newblock {\slshape \bibinfo{journal}{Software and Systems Modeling}}, pp.
  \bibinfo{pages}{1--25}, \doi{10.1007/s10270-022-01060-7}.

\bibitemdeclare{incollection}{moore1956gedanken}
\bibitem{moore1956gedanken}
\bibinfo{author}{E.F. \surnamestart Moore\surnameend} (\bibinfo{year}{1956}):
  \emph{\bibinfo{title}{Gedanken-Experiments on Sequential Machines}}.
\newblock In \bibinfo{editor}{Claude \surnamestart Shannon\surnameend} \&
  \bibinfo{editor}{John \surnamestart McCarthy\surnameend}, editors: {\slshape
  \bibinfo{booktitle}{Automata Studies}}, \bibinfo{publisher}{Princeton
  University Press}, \bibinfo{address}{Princeton, NJ}, pp.
  \bibinfo{pages}{129--153}, \doi{10.1515/9781400882618-006}.

\bibitemdeclare{article}{paige1987three}
\bibitem{paige1987three}
\bibinfo{author}{R.~\surnamestart Paige\surnameend} \& \bibinfo{author}{R.~E.
  \surnamestart Tarjan\surnameend} (\bibinfo{year}{1987}):
  \emph{\bibinfo{title}{Three partition refinement algorithms}}.
\newblock {\slshape \bibinfo{journal}{SIAM Journal on Computing}}
  \bibinfo{volume}{16}(\bibinfo{number}{6}), pp. \bibinfo{pages}{973--989},
  \doi{10.1137/0216062}.

\bibitemdeclare{article}{powerset}
\bibitem{powerset}
\bibinfo{author}{M.O. \surnamestart Rabin\surnameend} \&
  \bibinfo{author}{D.~\surnamestart Scott\surnameend} (\bibinfo{year}{1959}):
  \emph{\bibinfo{title}{Finite automata and their decision problems}}.
\newblock {\slshape \bibinfo{journal}{IBM Journal of Research and Development}}
  \bibinfo{volume}{3}(\bibinfo{number}{2}), pp. \bibinfo{pages}{114--125},
  \doi{10.1147/rd.32.0114}.

\bibitemdeclare{inproceedings}{ravikumar1996paralleldfa}
\bibitem{ravikumar1996paralleldfa}
\bibinfo{author}{B.~\surnamestart Ravikumar\surnameend} \&
  \bibinfo{author}{X.~\surnamestart Xiong\surnameend} (\bibinfo{year}{1996}):
  \emph{\bibinfo{title}{A Parallel Algorithm for Minimization of Finite
  Automata}}.
\newblock In: {\slshape \bibinfo{booktitle}{Proceedings of the 10th
  International Parallel Processing Symposium}}, \bibinfo{series}{IPPS '96},
  \bibinfo{publisher}{IEEE Computer Society}, \bibinfo{address}{USA}, pp.
  \bibinfo{pages}{187--191}, \doi{10.1109/IPPS.1996.508056}.

\bibitemdeclare{article}{stockmeyer1984simulation}
\bibitem{stockmeyer1984simulation}
\bibinfo{author}{L.~\surnamestart Stockmeyer\surnameend} \&
  \bibinfo{author}{U.~\surnamestart Vishkin\surnameend} (\bibinfo{year}{1984}):
  \emph{\bibinfo{title}{Simulation of parallel random access machines by
  circuits}}.
\newblock {\slshape \bibinfo{journal}{SIAM Journal on Computing}}
  \bibinfo{volume}{13}(\bibinfo{number}{2}), pp. \bibinfo{pages}{409--422},
  \doi{10.1137/0213027}.

\bibitemdeclare{inproceedings}{tewari2002paralleldfa}
\bibitem{tewari2002paralleldfa}
\bibinfo{author}{A.~\surnamestart Tewari\surnameend},
  \bibinfo{author}{U.~\surnamestart Srivastava\surnameend} \&
  \bibinfo{author}{P.~\surnamestart Gupta\surnameend} (\bibinfo{year}{2002}):
  \emph{\bibinfo{title}{A Parallel {DFA} Minimization Algorithm}}.
\newblock In: {\slshape \bibinfo{booktitle}{Proc.\ of HiPC}}, {\slshape
  \bibinfo{series}{LNCS}} \bibinfo{volume}{2552},
  \bibinfo{publisher}{Springer}, pp. \bibinfo{pages}{34--40},
  \doi{10.1007/3-540-36265-7_4}.

\bibitemdeclare{book}{trachtenbrot1973finite}
\bibitem{trachtenbrot1973finite}
\bibinfo{author}{B.A. \surnamestart Trakhtenbrot\surnameend} \&
  \bibinfo{author}{J.M. \surnamestart Barzdin\surnameend}
  (\bibinfo{year}{1973}): \emph{\bibinfo{title}{Finite automata: behavior and
  synthesis}}.
\newblock \bibinfo{publisher}{North-Holland Publishing}.

\bibitemdeclare{inproceedings}{ullman1990high}
\bibitem{ullman1990high}
\bibinfo{author}{J.~\surnamestart Ullman\surnameend} \&
  \bibinfo{author}{M.~\surnamestart Yannakakis\surnameend}
  (\bibinfo{year}{1990}): \emph{\bibinfo{title}{High-Probability Parallel
  Transitive Closure Algorithms}}.
\newblock In: {\slshape \bibinfo{booktitle}{Proc.\ of SPAA 1990}}, pp.
  \bibinfo{pages}{200--209}, \doi{10.1145/97444.97686}.

\bibitemdeclare{incollection}{wijs2015}
\bibitem{wijs2015}
\bibinfo{author}{A.J. \surnamestart Wijs\surnameend} (\bibinfo{year}{2015}):
  \emph{\bibinfo{title}{{GPU} Accelerated Strong and Branching Bisimilarity
  Checking}}.
\newblock In \bibinfo{editor}{C.~\surnamestart Baier\surnameend} \&
  \bibinfo{editor}{C.~\surnamestart Tinelli\surnameend}, editors: {\slshape
  \bibinfo{booktitle}{Proc.\ of TACAS}}, {\slshape \bibinfo{series}{LNCS}}
  \bibinfo{volume}{9035}, \bibinfo{publisher}{Springer}, pp.
  \bibinfo{pages}{368--383}, \doi{10.1007/978-3-662-46681-0_29}.

\end{thebibliography}
